\documentclass[11pt]{article}
\usepackage{amsmath}
\usepackage{amssymb}
\usepackage{graphicx}
\usepackage{color}
\usepackage{amsthm}
\usepackage{amsfonts}
\usepackage{amscd}
\usepackage{mathrsfs}
\usepackage{bm}
\usepackage{enumerate}
\usepackage{algorithmic, algorithm}

\newcommand{\bbR}{\mathbb{R}}

\newcommand{\bx}{\bm{x}}
\newcommand{\by}{\bm{y}}

\newcommand{\cM}{\mathcal{M}}

\newcommand{\p}{\partial}

\newcommand{\Rd}{\bbR^d}

\begin{document}
\title{Deep Residual Learning and PDEs on Manifold}
\date{}

\author{
Zhen Li %
\thanks{Department of Mathematics, Hong Kong University of Science \& Technology, Hong Kong. \textit{Email: mazli@ust.hk.}}%
\and
Zuoqiang Shi%
\thanks{Yau Mathematical Sciences Center, Tsinghua University, Beijing, China,
100084. \textit{Email: zqshi@tsinghua.edu.cn.}%
}
}
\maketitle
\graphicspath{{pics/}}


\begin{abstract}
\noindent
In this paper, we formulate the deep residual network (ResNet) as a control problem of transport equation. In ResNet, the transport equation is
solved along the characteristics. Based on this observation, deep neural network is closely related to the control problem of PDEs on manifold.
We propose several models based on transport equation, Hamilton-Jacobi equation and Fokker-Planck equation. The discretization of these PDEs on point cloud
is also discussed.
\end{abstract}

\noindent
\textbf{keywords:} Deep residual network; control problem; manifold learning; point cloud; transport equation; Hamilton-Jacobi equation

\section{Deep Residual Network}
\begin{figure}[H]
\centering
\begin{tabular}{c}
\includegraphics[width=0.1\textwidth]{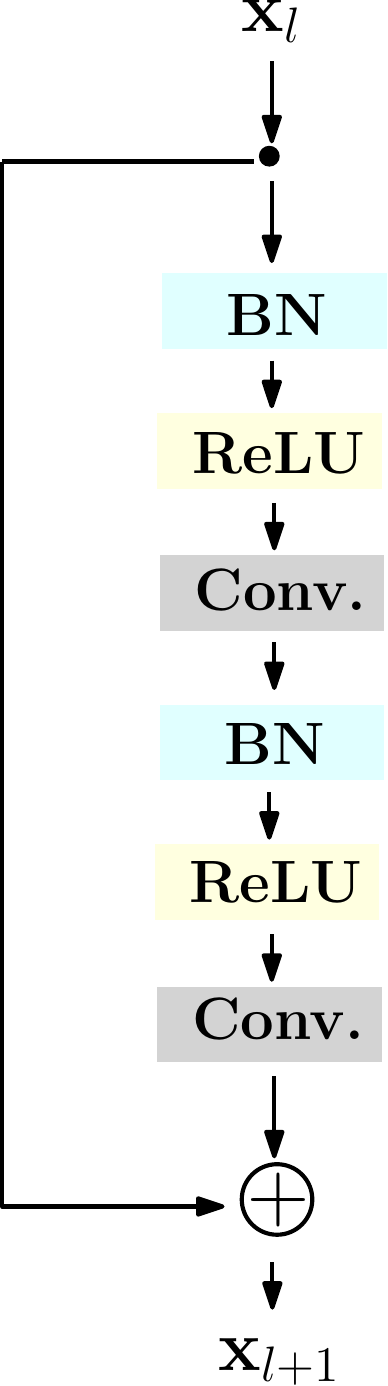}
\end{tabular}
\caption{Building block of residual learning \cite{resnet-2}.}\label{fig:residual}
\end{figure}

Deep convolution neural networks have achieved great successes in image classification.
Recently, an approach of deep residual learning is proposed to tackle the degradation in the classical deep neural network \cite{he2016deep,resnet-2}.
The deep residual network can be realized by adding shortcut connections in the classical CNN.
A building block is shown in Fig. \ref{fig:residual}. Formally,  a building block is defined as:
$$\by = \mathcal{F}\left(\bx, \{\bm{W}_i\}\right) + \bx.$$
Here $\bx$ and $\by$ are the input and output vectors of the layers.
The function $\mathcal{F}\left(\bx, \{\bm{W}_i\}\right)$ represents the
residual mapping to be learned. In Fig. \ref{fig:residual},
$\mathcal{F} = \bm{W}_2 \cdot \bm{\sigma}(\bm{W}_1 \cdot\bm{\sigma}(\bx))$ in which $\bm{\sigma}=\text{ReLU}\circ \text{BN}$
denotes composition of ReLU and Batch-Normalization.


\section{Transport Equation and ResNet}

Consider the the terminal value problem of linear transport equation in $\mathbb{R}^d$:
\begin{align}                                                       \label{eq_transport}
\left\{
\begin{aligned}
\frac{\p u}{\p t} - \bm{v}(\bx,t)\cdot \nabla u = 0, &\quad \bx\in \Rd, t \ge 0,\\
u(\bx,1) = f(\bx), &\quad \bx\in \Rd.
\end{aligned}
\right.
\end{align}
where $\mathbf{v}(\bx, t)$ is a given velocity field, 
$f$ is the composition of the output function and the fully connected layer. If we use the softmax activation function, 
\begin{equation}
\label{terminal}
f(\bx)={\rm \bf softmax}(\bm{W}_{FC}\cdot\bx),
\end{equation}
where $\bm{W}_{FC}$ is the weight in the fully connected layer, softmax function is given by
$${\rm \bf softmax}(\bx)_i=\frac{\exp(x_i)}{\sum_j \exp(x_j)}.
$$ 
which models posterior probability of the instance belonging to each class.

It is well-known that the solution at $t=0$ can be approximately solved along the characteristics:
\begin{equation}
\label{Characteristic-Eq}
\frac{d\bx(t;\bx)}{dt}=\mathbf{v}\left(\bx(t;\bx), t\right),\quad \bx(0;\bx)=\bx.
\end{equation}
We know that along the characteristics, $\mathbf{u}$ is a constant:
$$
u(\bx, 0) = u\left(\bx(1;\bx), T\right)=f(\bx(1;\bx)).$$
Let $\{t_k\}_{k=0}^L$ with $t_0=0$ and $t_L=1$ be a partition of $[0, 1]$. The characteristic of the transport equation (\ref{Characteristic-Eq}) can be solved by using simple forward Euler discretization from $\bx^0(\bx)=\bx$:
\begin{equation}
\label{Dis-Characteristic}
\mathbf{X}^{k+1}(\bx) = \mathbf{X}^k(\bx) + \Delta t \mathbf{v}(\mathbf{X}^k(\bx), t_k),
\end{equation}
where $\Delta t$ is the time step. If we choose the velocity field such that 
\begin{equation}
\label{velocity}
\Delta t\,\bm{v}(\bx, t) =\bm{W}^{(2)}(t)\cdot \bm{\sigma}\left(\bm{W}^{(1)}(t)\cdot \bm{\sigma}(\bx)\right),
\end{equation}
where $\bm{W}^{(1)}(t)$ and $\bm{W}^{(2)}(t)$ corresponds to the 'weight' layers in the residual block, $\sigma={\rm ReLU}\circ {\rm BN}$, 
one step in the forward Euler discretization \eqref{Dis-Characteristic} actually recovers one layer in the deep ResNets, Fig. \ref{fig:residual}. 
Then the numerical solution of the transport equation \eqref{eq_transport} at $t=0$ is given by
\begin{equation}
\label{output-transport}
u(\bx,0)=f(\mathbf{X}^L(\bx))
\end{equation}
This is exactly the output we get from the ResNets.

If $\bx$ is a point from the training set, we already have a labeled value $g(\bx)$ on it. Then we want to match the output value given in \eqref{output-transport} and $g(\bx)$ to train the parameters in the velocity filed \eqref{velocity} and the terminal value.

Based on above analysis, we see that the training process of ResNet can be formulated as an control problem of a transport equation in $\mathbb{R}^d$.
\begin{align}                                                       \label{eq_transport_inverse}
\left\{
\begin{aligned}
\frac{\p u}{\p t}(\bx,t) - \bm{v}(\bx,t)\cdot\nabla u(\bx,t) = 0, &\quad \bx\in \Rd, t \ge 0,\\
u(\bx,1) = f(\bx), &\quad \bx\in \Rd,\\
u(\bx_i,0)=g(\bx_i),&\quad \bx_i\in T.
\end{aligned}
\right.
\end{align}
where $T$ denotes the training set. $g(\bx_i)$ is the labeled value at sample $\bx_i$. Function $u$ may be scalar or vector value in different applications.

So far, we just formulate ResNet as a control problem of transport equation. This model will inspire us to get new models by replacing different component in the control problem.


\section{Modified PDE model}
From the PDE point of view, ResNet consists of five components:
\begin{itemize}
\item PDE: transport equation;
\item Numerical method: characteristic+forward Euler;
\item Velocity filed model: $\bm{v}(\bx,t) = \bm{W}^{(2)}(t)\cdot \bm{\sigma}\left(\bm{W}^{(1)}(t)\cdot \bm{\sigma}(\bx) \right)$;
\item Terminal value: $f(\bx)=\textbf{softmax}(\bm{W}_{FC}\cdot\bx)$;
\item Initial value: label on the training set $g(\bx_i),\; \bx_i\in T$.
\end{itemize}

In five components listed above, the last one, "initial value", is given by the data and we have no other choice. All other four components, we can 
consider to replace them by other options. Recently, there are many works in replacing forward Euler by other ODE solver. 
Forward Euler is the simplest ODE solver. By replacing it to other solver, we may get different network. In some sense, in DenseNet \cite{densenet},
forward Euler is replaced by some multi-step scheme. There are many numerical schemes to solve the ODE \eqref{Characteristic-Eq}. 
Some numerical scheme may be too complicated to be used in DNN. Constructing good ODE solver in DNN is an interesting problem worth to exploit in our future work.  

\subsection{Terminal Value}

\begin{figure}[H]
\centering
\begin{tabular}{ccccccc}
\includegraphics[width=0.15\textwidth]{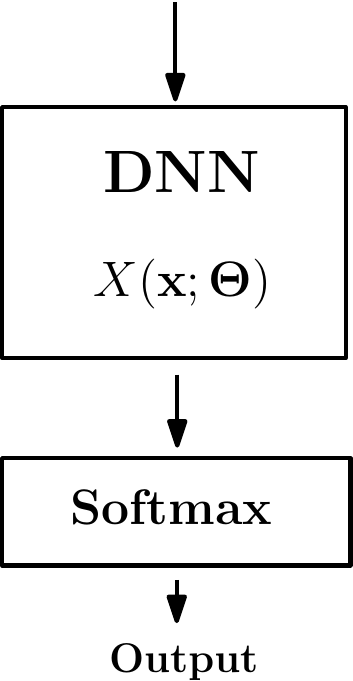}& &&
\includegraphics[width=0.15\textwidth]{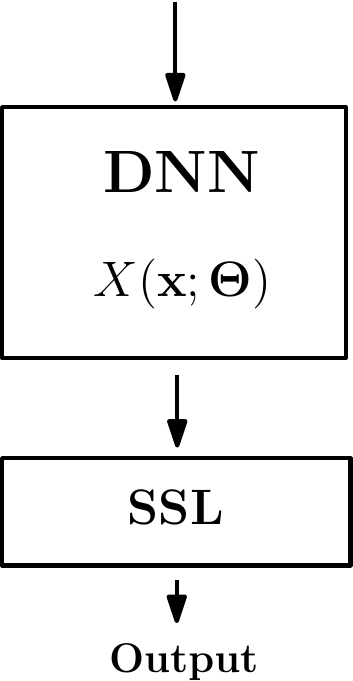}&&&
\includegraphics[width=0.15\textwidth]{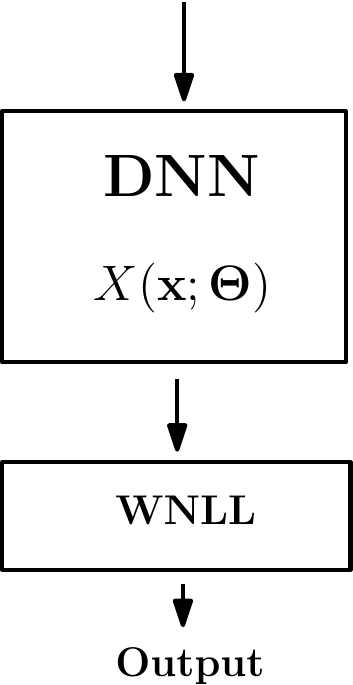}\\
(a)&&&(b)&&&(c)
\end{tabular}
\caption{(a): standard DNN; (b): DNN with semi-supervised learning; (c): DNN with last layer replaced by WNLL.}\label{fig:DNN_WNLL}
\end{figure}

From the control problem point of view, the softmax output function is not a good choice for terminal vaule, since they are pre-determined, may be far from the real value.
Semi-supevised learning (SSL) seems provides a good way to get terminal value instead of the pre-determined softmax function as shown in Fig. \ref{fig:DNN_WNLL}(b).  


Recently, we try to use weighted nonlocal Laplacian \cite{WGL} to replace softmax and the results is pretty encouraging \cite{deep-wnll}. 

\subsection{Velocity Field Model}

In  transport equation, we need to model a high dimensional velocity filed.
In general, it is very difficult to compute a high dimensional vector field. In the application associated to images, the successes of CNN and ResNet have proved that
the velocity filed model based on convolutional operators, \eqref{eq:vel-resnet}, is effective and powerful.
\begin{align}
\label{eq:vel-resnet}
\bm{v}(\bx,t) = \bm{W}^{(2)}(t)\cdot \bm{\sigma}\left(\bm{W}^{(1)}(t)\cdot \bm{\sigma}(\bx) \right).
\end{align} 
However, this is not the only way to model the velocity field. Moreover, for the applications in which convolutional operator makes no sense, we have to propose 
alternative velocity model. Here, we propose a model based on Hamilton-Jacobi equation to reduce the degree of freedom in the velocity field.

Notice that even though the velocity $\bm{v}(\bx,t)$ is a high dimensional vector field, in the tranport equation, only the component along $\nabla u$ is useful. 
Based on this observation, one idea is only model the component along $\nabla u$ by introducing $\bar{v}(\bx,t)=\bm{v}(\bx,t)\cdot \bm{n}(\bx,t)$ and
$\bm{n}(\bx,t)=\frac{\nabla_{\mathcal{M}} u}{|\nabla_{\mathcal{M}} u|}$. Then the transport equation in \eqref{eq_transport_inverse}
becomes a Hamilton-Jacobi equation. We get a control problem of Hamilton-Jacobi equation.
\begin{align}                                                       \label{eq_HJ_Rd}
\left\{
\begin{aligned}
\frac{\p u}{\p t} -  \bar{v}(\bx,t)\cdot
| \nabla u |&= 0, \quad \bx\in \Rd, t \ge 0,\\
u(\bx,0) &= f(\bx), \quad \bx\in \mathcal{M},\\
u(\bx_i,1)&=g(\bx_i),\quad \bx_i\in T.
\end{aligned}
\right.
\end{align}

In \eqref{eq_transport_inverse}, the velocity field, $\bm{v}(\bx,t)$, is a $d$-dimensional vector field.
Meanwhile, in the Hamilton-Jacobi model, we only need to model a scalar function $\bar{v}(\bx,t)$.
The number of parameters can be reduced tremendously. To model a scalar function is much easier than model a high dimensional vector field. 
There are already many ways to approximate $\bar{v}(\bx,t)$.
\begin{itemize}
\item[1.] The simplest way is to model $\bar{v}(\bx,t)$ as a linear function with respect to $\bx$, i.e.
  \begin{equation}
    \label{eq:v-linear-HJ}
    \bar{v}(\bx,t)=\bm{w}(t)\cdot \bx+b(t),\quad \bm{w}(t)\in \mathbb{R}^d,\; b(t)\in \mathbb{R}.
  \end{equation}
Although $\bar{v}(\bx,t)$ is a linear function to $\bx$, the whole model \eqref{eq_HJ_Rd} is not a linear model since Hamilton-Jacobi equation is a nonlinear equation. 

\item[2.] Neural network is consider to be an effective way to approximate scalar function in high dimensioanl space. One option is a simple MLP with one hidden layer as shown in Fig. \ref{fig:v-mlp}.
\begin{figure}[H]
\centering
\begin{tabular}{c}
\includegraphics[width=0.6\textwidth]{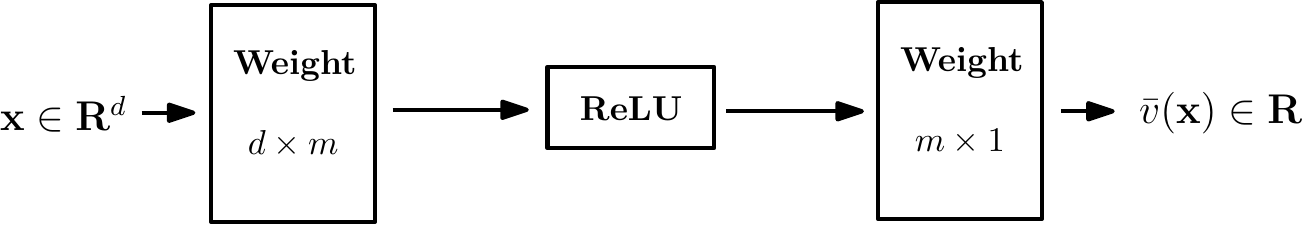}
\end{tabular}
\caption{MLP model with one hidden layer for $\bar{v}(\bx,t)$ in \eqref{eq_HJ_Rd}.}\label{fig:v-mlp}
\end{figure}
There are many other neural networks in the literature to approximate scalar function in high dimensioanl space.

\item[3.] Radial basis function is another way to approximate $\bar{v}(\bx,t)$. Radial functions centered at each sample point are used as basis function to approximate 
$\bar{v}(\bx,t)$, i.e.
\begin{equation}
  \label{eq:v-radial}
\bar{v}(\bx,t)=\sum_{\bx_j\in P} c_j(t)R\left(\frac{|\bx-\bx_j|^2}{\sigma_j^2}\right)
\end{equation}
where $c_j(t),\; j=1,\cdots,|P|$ are coefficients of the basis function. One often used basis function is Gaussian function, $R(r)=\exp(-r)$. 
\end{itemize}

Another difficulty in solving model \eqref{eq_HJ_Rd} is efficient numerical solver of Hamilton-Jacobi equation in high dimensional space. 
Hamilton-Jacobi equation is a nonlinear equation which is more difficult to solve than linear transport equation in \eqref{eq_transport_inverse}. 
Recently, fast solver of Hamilton-Jacobi equation in high dimensional space attracts lots of attentions and 
many efficient methods have been developed \cite{UCLA-HJ-3,UCLA-HJ-4,UCLA-HJ-5,UCLA-HJ-1,UCLA-HJ-2,PDE-E}.

\subsection{PDEs on point cloud}
\label{sec:manifold_PDE}

Another thing we can consider to change is the characteristic method in solving the transport equation. There are many other numerical method for transport equation based on 
Eulerian grid. However, in these methods, we need to discretize the whole space $\mathbb{R}^d$ by Eulerian grid, which is impossible when $d$ is high. So 
in high dimensional space $\mathbb{R}^d$, characteristic method seems to be the only practical numerical method to solve the transport equation.
On the other hand, we only need to solve the 
transport equation in the dataset instead of the whole $\mathbb{R}^d$ space. 
Usually, we can assume that the data set sample a low dimensional manifold $\cM\subset \mathbb{R}^d$, then the PDE can be confined on this manifold. The manifold is sampled by 
the point cloud consists of the data set including the training set and the test set. Then the numerical methods in point cloud can be used to solve the PDE.

\paragraph{Transport equation}
In the manifold, the transport equation model \eqref{eq_transport_inverse} can be rewritten as follows:
\begin{align}                                                       \label{eq_transport_manifold}
\left\{
\begin{aligned}
\frac{\p u}{\p t} -  \bm{v}(\bx,t)
\cdot \nabla_{\mathcal{M}} u &= 0, \quad \bx\in \mathcal{M}\subset\Rd, t \ge 0,\\
u(\bx,0) &= f(\bx), \quad \bx\in \mathcal{M},\\
u(\bx_i,1)&=g(\bx_i),\quad \bx_i\in T.
\end{aligned}
\right.
\end{align}
$\nabla_\cM$ denotes the gradient on manifold $\cM$.
Let $X: V\subset \mathbb{R}^m\rightarrow \cM\subset\mathbb{R}^d$ be a local parametrization of $\cM$ and $\theta\in V$.
For any differentiable function $u:\cM\rightarrow \mathbb{R}$,
let $U(\theta)=f(X(\theta))$, define
\begin{eqnarray}
  \label{eq:diff-M}
  D_ku(X(\theta))=\sum_{i,j=1}^m g^{ij}(\theta)\frac{\p X_k}{\p\theta_i}(\theta)\frac{\p U}{\p\theta_j}(\theta),\quad k=1,\cdots,d.
\end{eqnarray}
where $(g^{ij})_{i,j=1,\cdots,m}=G^{-1}$ and $G(\theta)=(g_{ij})_{i,j=1,\cdots,m}$ is the first fundamental form which is defined by
\begin{eqnarray}
  \label{eq:remainn}
  g_{ij}(\theta)=\sum_{k=1}^d\frac{\p X_k}{\p\theta_i}(\theta)\frac{\p X_k}{\p\theta_j}(\theta),\quad i,j=1,\cdots,m.
\end{eqnarray}
$\nabla_\cM$ is defined as
\begin{align}
  \label{eq:grad-manifold}
  \nabla_{\mathcal{M}} u=(D_1 u, D_2 u,\cdots, D_du).
\end{align}
In \eqref{eq_transport_manifold}, the manifold $\cM$ is sampled by the data set $P$ and $P$ is a collection of
unstructured high dimensional points.
 Unlike the classical numerical methods which solve PDE on regular grids (or meshes), in this case, we need to discretize PDE on unstructured high
dimensional point cloud $P$. To handle this kind of problems,
recently, point integral method (PIM) was developed to solve PDE on point cloud.
In PIM, gradient on point cloud is approximate by an integral formula \cite{PIM-variable}.
\begin{align}
  \label{eq:grad-PIM}
 D_k u(\bx) = \frac{1}{w_\delta(\bx)}\int_\cM \left(u(\bx)-u(\by)\right) (x^k-y^k) R_\delta(\bx,\by) d \by.
\end{align}
where $w_\delta(\bx)=\int_{\cM}R_\delta(\bx,\by) d \by$,
\begin{align}
  R_\delta(\bx,\by)= R\left(\frac{\|\bx-\by\|^2}{\delta^2}\right).
\end{align}
 The kernel function $R(r): \mathbb{R}^+ \rightarrow \mathbb{R}^+ $ is assumed to be a $C^2$ function with compact support.

Corresponding discretization is
\begin{align}
  \label{eq:grad-PIM-dis}
 D_k u(\bx) = \frac{1}{\bar{w}_\delta(\bx)}\sum_{\by\in P} \left(u(\bx)-u(\by)\right) (x^k-y^k) R_\delta(\bx,\by) V(\by).
\end{align}
$\bar{w}_\delta(\bx)=\sum_{\by\in P}R_\delta(\bx,\by) V(\by)$ and $V(\by)$ is the volume weight of $\by$ depends on the distribution of the point cloud in the manifold.

%

\paragraph{Hamilton-Jacobi equation}

We can also confine the Hamilton-Jacobi equation in \eqref{eq_HJ_Rd} in the manifold.
\begin{align}                                                       \label{eq_HJ_manifold}
\left\{
\begin{aligned}
\frac{\p u}{\p t} -  \bar{v}(\bx,t)\cdot
| \nabla_{\mathcal{M}} u |&= 0, \quad \bx\in \mathcal{M}\subset\Rd, t \ge 0,\\
u(\bx,0) &= f(\bx), \quad \bx\in \mathcal{M},\\
u(\bx_i,1)&=g(\bx_i),\quad \bx_i\in T.
\end{aligned}
\right.
\end{align}

On the point cloud, one possible choice to discretize $| \nabla_{\mathcal{M}} u |$ is
\begin{align}
  \label{eq:pim-grad-l2}
  | \nabla_{\mathcal{M}} u(\bx) |=\left(\int_\cM w(\bx,\by)(u(\bx)-u(\by))^2 d \by \right)^{1/2}.
\end{align}
$\bar{v}(\bx,t)$ can be modeled in the way discussed in the previous section.



\paragraph{PDEs with dissipation}
We can also consider to add dissipation to stabilize the PDEs.
\begin{align}                                                       \label{eq_vis_HJ_manifold}
\left\{
\begin{aligned}
\frac{\p u}{\p t} -  \bar{v}(\bx,t)\cdot
| \nabla_{\mathcal{M}} u| &= \Delta_\cM u, \quad \bx\in \mathcal{M}\subset\Rd, t \ge 0,\\
u(\bx,0) &= f(\bx), \quad \bx\in \mathcal{M},\\
u(\bx_i,t)&=g(\bx_i), \quad\bx_i\in S,\\
u(\bx_i,1)&=g(\bx_i), \quad\bx_i\in T\backslash S,\\
\frac{\p u}{\p\bm{n}}(\bx,t)&=0, \quad \quad\quad\bx\in \p\cM.
\end{aligned}
\right.
\end{align}
In the model with dissipation, we need to add constraints, $u(\bx_i,t)=g(\bx_i), \;\bx_i\in S\subset T$ in a subset $S$ of
the training set $T$. Otherwise, the solution will be
too smooth to fit the data due to the viscosity. The choice of $S$ is an issue. The simplest way is to choose $S$ at random.
In some sense, the viscous term maintains the regularity of the solution and the convection term is used to fit the data.


On the point cloud, the Laplace-Beltrami operator along with the constraints $u(\bx_i,t)=g(\bx_i), \;\bx_i\in S\subset T$
can be discretized by the weighted nonlocal Laplacian \cite{WGL}.

\section{Discussion}

In this paper, we establish the connection between the deep residual network (ResNet) and the transport equation. ResNet can be formulated as
solving a control problem of transport equation along the characteristics. Based on this observation, we propose several PDE models on the
manifold sampled by the data set. We consider the control problem of transport equation, Hamilton-Jacobi equation and viscous Hamilton-Jacobi equation.

This is a very preliminary discussion on the relation between deep learning and PDEs. There are many important issues remaining unresolved including
the model of the velocity field, numerical solver of the control problem and so on. This is just the first step in exploring the relation between deep learning and
control problems of PDEs.

\bibliographystyle{abbrv}
\bibliography{reference}
\end{document}